\documentclass[10pt,=preprint]{aastex}
\usepackage{color}
\begin{document}

\title{Short GRBs Viewed  from Far Off Axis}
\author{David Eichler$^{*}$}
\altaffiltext{*}{Dept. of Physics, Ben-Gurion University, Be'er-Sheva 84105, Israel}

\begin{abstract}
  The recent radio observations (Mooley et al, 2018) of a superluminal radio afterglow  following GRB 170817A are interpreted in terms of a jet impacting a baryonic cloak, which is presumably the material caught at the front of the jet as the latter emerges from a denser ejected material. Assuming that we the observers are located   {at a viewing angle of $\sim 0.2$ radians from the emitting material (perhaps slightly more from jet axis)}, we suggest that the Lorentz factor of the jet is  $\lesssim 20$ at the time of the prompt emission, and that, as suggested previously, it is accelerated to much higher values before finally decelerating during the afterglow phase. A less extreme example of a short GRB being observed off axis may have been GRB 150101b (Fong, et al., 2016). A feature of  GRBs viewed from large offset angles is a large afterglow isotropic equivalent energy as compared to prompt emission, as predicted (Eichler, 2017), and this is born out by the observations of these two GRB.
  
  It is also shown that the prompt emission of GRB 170817A, if seen way off-axis ($\theta \gg 1/\Gamma $), could not be made by internal shocks in the baryonic material that powers the afterglow. 
 
\end{abstract}

\section{Introduction }

Mooley et al  (2018) have recently measured apparent superluminal motion in the radio afterglow  emission of GRB 170817A, which was associated with a neutron star merger. The apparent motion is of order 4c  around the time of radio maximum ({\it circa} day 150 after the prompt emission). From this the authors infer that at the epoch of this maximum, the Lorentz factor of the radio emitting region is about (or slightly greater than) 4 and that it is seen from  an angle of about  0.25 to 0.35  radian, or about 14  to 20 degrees, from  the jet axis.  From the sharpness of the peak, they also infer that the angular width of the jet itself cannot be much more than about 0.1 radian  Hence, if the jet is axisymmetric, the distance to the closest part of its perimeter cannot be less  than about 0.15 radian from our  line of sight, and is more likely offset by at least 0.175 or 0.2 radians.  { Mooley et al. (2018) also conclude that the energy in the blast that make the afterglow has an isotropic equivalent energy $E_{iso} \equiv 10^{52}E_{52}$ ergs of about $10^{52}$ ergs . If the asymptotic Lorentz factor $\Gamma_a$ of the baryonic material that drives the blast is $10^{2.5}\Gamma_{a,2.5}$, then its isotropic equivalent mass $m_{iso} \equiv 10^{28.5}m_{28.5}$ g is then  $m_{iso} \simeq 10^{28.5} E_{52}/\Gamma_{a,2.5}$ g.

The angular extent of the material  that last interacted with ( i.e. emitted or scattered) the prompt emission need not be the same as that of the material driving the late time afterglow. In particular, the distribution of the former to  may be wider than the latter because the criterion for contributing to the prompt emission is determined by optical depth, which need not require much of the total mass,  whereas the contribution to late time afterglow depends on the angular distribution of ejected mass and momentum. Virtually all of the discussion below is qualified by this uncertainty. For the sake of concreteness however,  it is assumed below for concreteness that the observer's line of sight is 0.2 radians from the material that most contributes to the prompt emission and the latter material is taken to be a pencil beam that is somewhat closer to our line of sight than the jet axis. 

The relatively large viewing angle inferred for GRB 170817A was anticipated, (e.g. Eichler, 2017)   if only because the accompanying gravitational wave signal selected it for unusually close proximity while more distant GRBs are selected for small viewing angles, where relativistic beaming makes them bright enough to be detected. The  observed initial rise in afterglow intensity with time, which has been recorded  since the discovery, is consistent with  an off-axis jet, whose forward shock becomes increasingly visible as it decelerates (e.g. Lazzati et al, 2018; Lyman et al, 2018; Zhang, et al. 2018; Lamb and Kobayashi, 2018;; Ioka, and Nakamura, 2018; Resmi  et al. 2018).  However, the relatively sharp reversal from rising to declining afterglow in both the radio and X-ray (Alexander et al. 2018; Margutti et al., 2018; Lamb, Mandel, and Resmi, 2018), together with the apparently  superluminal motion in the radio, seems to clinch the case for a strongly relativistic off-axis jet over a quasi-spherical, mildly relativistic outflow.}

That short GRB {as a class} are  viewed from a larger viewing angle than long ones has been advocated for many years,
 [e.g. Eichler, Guetta, and Manis (2009)].  {In this particular model, photons (or in any case a baryon poor fireball) overtake   baryonic matter in front of them, scattering off it and accelerating it. The baryonic matter so accelerated eventually powers the afterglow. }  While a wider viewing angle is kinematically associated with lower Doppler factors and hence longer observed durations,  in this particular model, photons seen near the peak of the burst are last scattered (or emitted and then never scattered) from an accelerating surface that is, at the time of the last scattering, moving at a Lorentz factor $\Gamma$ of at least $1/\theta$ relative to the observer's line of sight.  After the peak,  the Lorentz factor is larger, causing the observer to see  a ''soft tail'' (often called ``extended emission'') of softening, dimming emission. { This tail could also be due in part to a light echo off parts of the jet that are oriented further away from our line of sight.} The short duration at larger viewing angle is attributed to the shorter acceleration time of the surface of last scattering when $\theta$ is large and $\Gamma_{peak}$, therefore, small.
{ The acceleration time (in the frame of the central engine) $d (ln{1-\beta})/dt$ is of order $\Gamma^3 m_{iso} c/L_{iso}$ s (Eichler and Manis, 2007), where $L_{iso}\equiv 10^{53}L_{53}$ is the isotropic equivalent luminosity along the jet axis,   which, for $m_{28.5}$, $E_{52}$ and $\Gamma_{a, 2.5}$ all  of order unity, is of order $10^{-3.5} \Gamma^3/L_{53}$ s. In observer time, this time interval is compressed by reduction in propagation distance by the factor $ (1-\beta  \cos \theta)$, which, for GRB 170817A, is estimated below to be a factor of $(1- \cos\theta ) \simeq 0.02$.  Altogether, we might expect  that within the first 0.1 seconds of observer time, the baryons are accelerated to a  Lorentz factor of $\sim 25 L_{53}^{1/3}$.}
 
   In anticipation of the announcement of the discovery of GRB 170817A, and in more detail just afterwards, it was noted (e.g. Eichler, 2017) that this GRB, because it was so close, could be seen at even larger viewing angles, where it would  appear softer and fainter than most GRB,  yet where it would be within a {broader and therefore}  more likely range of viewing angles  for a random distribution of observers.  As the spectral peak  - at 180 keV +,- 60 keV - was about 22 times softer than the spectral peak of GRB 090510 (which was taken as the benchmark spectral peak of an on-axis observer), it was conjectured  in the above reference that the viewing angle was about $\sqrt{21}/\Gamma \sim 4.5/\Gamma$.  This estimate follows from the fact that the observed photon energy $E''$ depends on the viewing angle $\theta$ to the direction of motion of the scatterer $\hat \beta$ as $(1-\beta)E/(1-\beta \cos \theta) \sim E/[1+(\theta \Gamma)^2 \sim E/22]$  where E is the energy seen by an on axis observer given a monochromatic, monodirectional beam that overtakes the scatterer from behind. That is, E is the photon energy in the observer frame prior to scattering. 
 
 Accordingly, the luminosity, which from a pencil beam scales as $[1/\Gamma(1-\beta \cos \theta)]^4$ would scale as $(E''/E)^4$, which, in the case of GRB 0170817A, would be $\sim 22^{-4} \sim 4 \cdot 10^{-6}$. Remarkably, this gives a reasonable estimate of the peak luminosity of GRB 170817A, which was of order $2.5 \cdot 10^{47}$ erg s$^{-1}$, relative to the peak luminosity of GRB 090510 ($L_{iso} \simeq1 \cdot 10^{53}$ erg s$^{-1})$,  a short GRB that we have previously argued was seen on-axis (Eichler, 2017; Eichler 2014), and sets the standard expectation for an on-axis observer.

 \section{Significance of Afterglow Observations}

 Most remarkably, however, the radio afterglow observations support the conclusion  that there is fact was a more or less normal isotropic equivalent energy output for a short GRB, $E_{iso}\sim 10^{52}$ ergs, beamed away from the observer.\footnote{This fact had been challenged earlier by advocates  (Mooley et al. 2017) of a ``choked GRB'' interpretation of the low luminosity.}  
 One might suspect, therefore, that the prompt emission is viewed from the same angle as the radio emission, $\gtrsim 0.2$ radians. If the Lorentz factor $\Gamma$ of the scatterer (i.e. the surface of last interaction of the prompt $\gamma$-ray emission)  obeys  $\Gamma \lesssim 25$, then $\theta$ would indeed be $\gtrsim 4.5/\Gamma$, adequately accounting for the subluminous character of this  particular short GRB. 
 
   GRB170817A has much in common with GRB150101B (Fong et al, 2016), also believed to have been viewed from a wide viewing angle, in that the afterglow fluence is much higher than that of the prompt emission.  Both GRB support the notion that most of the acceleration of the baryonic material that powers the afterglow took place {\it after} the  prompt emission phase.  This fits the picture (Eichler and Manis, 2008; Eichler, Guetta, Manis 2009; Eichler, 2017) that the prompt emission in our direction is aborted by the acceleration of the scattering material to values much greater than $1/\theta$. It  connects the  unusual brevity of GRB 150101B ($T_{90}$ of only 18 ms) to the unusually high ratio of afterglow fluence to prompt fluence, as both correlate with large $\theta$.\footnote{GRB 170817A is also unusually short, even among short GRB (Kaneko, et al, 2015) , if the spike, as defined by Kaneko et al (2015) is taken to be the 100 ms peak, and the soft, extended emission is taken to be 2 seconds.}
 
 Now  Veres et al (2018) have recently argued that in fact the spectral peak may be somewhat larger at peak luminosity, which would mean, in the context of Eichler (2017),  that  a smaller value $E/E''$ should be used, perhaps only 10 rather than 22, and that $\theta$ may be a small as $3/\Gamma$, giving a smaller reduction of  apparent luminosity due to off-axis kinematic effects.\footnote{They also assume that the ratio of prompt  emission on and off- axis varies as $(E/E'')^2$, which is inappropriate for an observer who sees the jet as a pencil beam. They also equate the afterglow energy with the on-axis prompt emission, and this is also without justification.   Each of these {unjustified assumptions} lead to an overestimate of $E/E''$.}  However, given the uncertainties in the isotropic equivalent luminosity seen by an on-axis observer, this is not a serious concern. The principal uncertainty is that the vast majority of the kinetic energy that powers the afterglow may have entered the scatterer {\it after} the peak of the prompt emission, as discussed in Eichler (2017).  This is supported by the wide range in the ratio r of afterglow fluence ${\cal F}_{afterglow}$ to prompt fluence ${\cal F}_{prompt}$ over short GRB, ranging from $r\simeq  10^3$ in the cases of highly off-axis viewing angles such as GRB150101B (Fong et al, 2016) down to $r \simeq 10^{-4}$ in the case of GRB090510 which was  probably viewed nearly face on.  Additional uncertainty  can arise from intrinsic scatter in r, apart from viewing angle considerations, especially because nearby GRB are more likely to be at the low end of this scatter, while distant ones are at the high end. Yet more uncertainty arises in the  coverage factor of the scattering material ahead of the fireball.  
 
 In any case, there now seems to be agreement among theorists that  the GRB 170817A was observed way off axis and that the prompt fluence was well below average for a short GRB, but that it was nevertheless accompanied by a long term afterglow of average fluence. As such, it resembles the GRB 150101B, which had a low fluence (though average luminosity) and a huge $\sim 10^3$ afterglow to prompt fluence ratio. Probably GRB 170817A is the even more extreme case as it is viewed even further off axis.
 
 \section{The Transparency Issue}
 
 Even within the framework of an off-axis kinematics picture, there may still be open questions. For example there is the question of whether photons that are scattered off-axis  (i.e. in the backward hemisphere in the frame of the scatterer) make any further interactions with not-yet scattered ones, or more generally whether pairs are reestablished by the interaction of photons with baryonic material. This depends on the details.
   
  There are several cases that need to be considered. 
  
  \noindent a) a baryonic shell impacted from behind by an ultrarelativistic jet. The jet is assumed to be mostly or all photons that peak at several MeV.  In the case  of GRB 090510, the spectral peak of the photospheric emission was  $E_p$ was at  $E_p \simeq 4$ MeV. A photon observed at earth of energy $E''$ had an energy $E'$ in the scatterer frame of $E'_o= E''\Gamma(1-\beta \cos \theta)$, whereas the photons in the not yet scattered jet have an energy $E'_j = E/\Gamma(1+\beta)$. The criterion for pair production $E'_oE'_j \ge (2m_ec^2)^2$ is satisfied when $E''E (1-\beta \cos \theta)/(1+\beta) \simeq  E''E (1-\beta \cos \theta)/2\ge (2m_e c^2)^2$. 
  For $1-\beta \cos \theta =0.02$, the case of GRB 170817A with $\theta$ taken to be 0.2,  the criterion for pair production is that 
  
  \begin{equation} 
  E''E \ge 400(m_e c^2)^2 \sim 200 MeV m_e c^2. 
  \end{equation}
   So for example if a photon is observed at $E''= 512$ MeV, then it could have pair produced only with a  photon that would have been observed at $E \ge 200$ MeV by an  on-axis observer.   So even if we assume that the photons in the jet had a Comptonization spectrum extending all the way to  200 MeV, if the temperature of those photons is only $2$ MeV, than the fraction of photons  that could  pair produce would be proportional to the Boltzmann factor $e^{-100}$ which would imply a negligible amount of pairs produced.\footnote{We neglect the possibility of a Compton tail of scattered photons extending in  energy well beyond the range of the observed photons, because the Compton recoil off the scattering material would soften any such  tail. There would be no photons much above $E' \sim mc^2$. Moreover, any pairs that are created are kept out of the body of the jet by radiation pressure. }

 \noindent b) the case where photons are Comptonized with a temperature $T'=E'_p/2$.  Here we need to worry about photons at higher energies than the observed ones pair producing with other unobserved photons, because such pairs would block even photons below the pair production threshold.  Given the photon isotropic equivalent luminosity $10^{53}L_{iso,53}$ erg s$^{-1}$ of the jet,  and the radius of emission, which is taken to be  $R=c\delta t''/(1-\cos \theta)=1.5 \cdot 10^{11}$ cm,  we compute the optical depth to pairs in Compton equilibrium with the photons of temperature T', given a comoving energy density of 
 
 \begin{equation}
 U' =  10^{53}L_{iso,53}{{(1-\beta)}\over{(1+\beta)}}/\left[4\pi c R^2\right]
 \end{equation}
 The ratio of pair density to photon energy density is (Svensson, 1984)
 \begin{equation}
 n'/U'=1.2 [8/\pi]^{1/2}[1 + 0.372 y^{-1/2} 
+ 0.472 y^{-1} +(3/2 \pi)^{1/2} 1.2y^{-3/2}]^{-1} y^{-3/2} exp[-y]/3kT' \equiv \eta  y^{-3/2} e^{-y}/3kT'
 \end{equation}
   where $y\equiv m_e c^2/kT'$, and where the numerical factor $\eta$ will turn out to be $\sim 1.7$.
   So 
   \begin{equation}
\tau =\left[ 10^{53}L_{iso,53}{{(1-\beta)}\over{(1+\beta)}}/12\pi kT' c R\right]\eta y^{-3/2}e^{-y}\sigma_T /\Gamma 
\label{transp2}
 \end{equation}
 
 The radius of emission R is constrained by the observation that the delay between the gravitational signal and the  GRB of 1.7 s. The quantity $R(1-\cos \theta)/c=0.02R/c $ must  therefore be less than 1.7s, so $R\le 2.5 \cdot 10^{12}$ cm.  Another constraint comes from the fact that the peak of the emission is only $\sim 0.1$ s, so even if the material emits a pulse of locally infinitesimal duration in our direction the delay between the arrival of the closest and furthest region of the last interaction with baryons, $2R\theta_o (1- \cos \theta)/c$, must be less that 0.1 s, implying that  $R\le 1.5 \cdot 10^{11}/\theta_o$ s. Here $\theta_o$ is the opening angle of the jet and may be as small as 0.05. Altogether, 
     $R$ may be at most of order $10^{12}$, and the condition that $\tau$ be less than $\sim 10$  implies that $T'\lesssim 35 $ KeV, and $E_p' = 2 kT' \lesssim 70$ KeV.{\bf \footnote{We choose an upper limit of 10 on $\tau$;  $1\le \tau \le 10$ is admissible because the low luminosity of GRB 170817A allows an optical depth of up to 10.}} If $1-\beta \cos\theta$ can be approximated as $1-\cos \theta \simeq \theta^2/2$, then  the observation of a peak frequency $E''$ of about 500 KeV in the observer frame implies 
     
     \begin{equation}
     E_p' = E_p'' \Gamma(1-\beta \cos \theta) =0.02 \Gamma \cdot 500 \rm KeV  \le 70 \rm KeV;
     \end{equation}
     whence $\Gamma \lesssim 7$.
     It is emphasized that this constraint is contingent on the assumption that  the pairs and photons are in Compton equilibrium. The observation  that the spectrum of GRB 170817A has a soft component below  the  peak casts doubt on whether this is a good assumption. {Moreover, the peak energy $E_p''$ is extremely uncertain (Veres et al, 2018); if it is only 300 KeV, well within the error bars of Veres et al,  then $\Gamma$ is constrained more weakly to be  $\lesssim 12$. So  the assumption that the photons are in Compton equilibrium with a  trace residuum of pairs, though not necessarily the case,  would be more constraining; it is just marginally  compatible with the rest of the model.\footnote{That the plasma is only marginally transparent at the beginning of the GRB would be attributable to opacity contributing to the delay of the  appearance of the GRB following the GW event. } }
     
      \noindent c) the baryons that power the afterglow are already mixed with the on-axis photons at the time of the prompt emission:  In this case the afterglow itself establishes the baryon environment through which those photons must escape.   The density of baryons, if they power  afterglow  with about the same luminosity as the  prompt emission that would be seen by an on-axis observer, is established by the condition that they must contain enough energy:
      
  \begin{equation}
 U  =  L _{iso}  /\left[4\pi c R^2\right] =  \Gamma n m_p c^2 
 \end{equation}
 and
   \begin{equation}
  \tau  = n'\sigma_T R/\Gamma  = n\sigma_T R/\Gamma^2  = \left[L_{iso}  \sigma_T /4\pi   Rm_p c^2 \Gamma^3 \right]= 2.2 \cdot 10^8 L_{iso,53}R_{12}^{-1} \Gamma^{-3}
 \end{equation}
 so that  $\tau\lesssim 1$ implies $\Gamma \gtrsim 600 L_{iso,53}^{1/3} R_{12}^{-1/3}$. In other words, if the observed gamma rays had needed to escape from within the baryons that power the afterglow,  then they could do so only if the latter had a Lorentz factor  of several hundred, and this would be too  ultrarelativistic to give significant amounts of off-axis radiation. This lower limit on $\Gamma$ is well known in the GRB literature. 
 
 We thus derive an important result: There cannot be much off-axis emission from within the material that powers the long-term  afterglow. It would be too opaque to emit far off axis. Photons generated from within the jet by internal shocks would be dragged away from the observer's line of sight by the motion and opacity of the baryonic fluid. The off-axis viewing hypothesis for GR 170817A works only if the photons impact the baryons from without and scatter off its surface.  We interpret this to mean that the baryons  are probably from  the merger ejecta and are plowed up by the jet  from behind.

    \section{Conclusions and Further Discussion}
   
   The  surprisingly transparent nature of GRBs, given their high compactness, is perhaps the central mystery of GRBs.  Two completely different explanations have been attempted: One is that huge Lorentz factors dilate both comoving time and distance scales  of  the fireball (Rees and Meszaros, 1992), the effect of which is to lower the density of the outgoing material as well as its true compactness required to account for rapid variability.  The other (Levinson and Eichler, 1993) invokes a baryon free corridor - probably connected to an event horizon -  that the photons are either produced inside of or scattered into. If they are scattered by swept up material at the leading edge nearly opposite to the direction of the scatterer's motion (i.e. back towards the central engine that the scatterer is presumed to be receding from)  then   they are observed at ``large'' viewing angles'' ( i.e. at least several times $1/\Gamma$) relative to the motion of the scatterer.  
   
   {A qualitative difference between the two accounts of GRBs is that one - using baryonic kinetic energy as the primary source of energy -  puts the energy in baryons which only later manage to generate photons via shocks, whereas the other   - a fireball that is nearly baryon free - has the energy flowing into baryons from an essentially non-baryonic fireball and predicts that that the baryons that eventually power the afterglow will be accelerated by the push of a non-baryonic fireball from which the prompt emission originates. 
   The baryon poor fireball scenario  thus lends itself to a scenario is which the length of prompt GRB emission may depend on the timescale over which the observer is within the $1/\Gamma$ cone of emission, which in turn depends on the viewing angle of the observer. The dichotomy between short and long GRB, and their different progenitors, in fact, may be that the range of viable viewing angles depends strongly on the progenitor - i.e. merging neutron stars (short GRB) allow viewing from a wider range of angles than collapsars within giant envelopes (long GRB).}
   
   
It should also be recalled that the combination of acceleration and viewing angle, with few free parameters,  has considerable predictive power and , unlike the alternative models, quantitatively explains the Amati relation  for long GRB (Eichler and Levinson 2004, 2006), the flat phase of afterglows (Eichler, 2005;  Eichler and Jontoff-Hutter, 2005)  spectral hard-to-soft evolution (Manis and Eichler 2007, 2008) and the differences between short and hard GRB (Eichler, Guetta and Manis, 2009, Eichler, 2017).  In particular, to accommodate the Amati relation extending down to several KeV and $E_{iso} \sim 10^{48}$ erg.s,  i.e. almost six orders of magnitude in isotropic equivalent energy, the viewing offset must be as large as $30/\Gamma$ for the softest sources.
    
    Afterglow observations of GRB 170817A provide  an unprecedented arena for testing hypothesis that the above features of GRB are due explainable by viewing angle kinematics, because the viewing angle relative to the jet axis is probably larger than for any other GRB to date, and because the large viewing angle is verifiable in several different ways (gravitational wave polarization, radio afterglow, $ etc.$).  In an off-axis viewing scenario, photons coming from the jet that end up in our direction must be traveling nearly backward in the frame of the jet material, and here we have carefully considered the question of whether they could get out through the stream of outwardly propagating photons from the central source without pair producing. We find that there is no reason they shouldn't, unless their energy is much higher than those detected in the NaI detectors of the GBM. 
    
       {Using the above considerations the following scenario for prompt emission can be suggested based on figure 6 of Goldstein et al: The GRB begins with $100 \lesssim E_p \lesssim 500 $ KeV, with $7 \lesssim \Gamma \lesssim 12$ and, by the end, where the spectrum peaks at  only 10 to 20 KeV, it has been accelerated to $20 \lesssim \Gamma \lesssim 40$.  Attributing the softening to acceleration alone would not account quantitatively for the brightness of the soft tail.  The soft tail  could be due to a light echo off parts of the baryonic shell whose orientation is further from the line of sight. A detailed account of the light curve, though beyond the scope of this paper, is important.} 
    
    We also find an {\it upper} limit on the Lorentz factor if the prompt emission is viewed from a large offset, because, for a given value of $(1-\beta \cos \theta)$,  the inverse Doppler factor $\Gamma(1-\beta \cos\theta)$, which relates the comoving photon energy (in the frame of the scatterer) to the observed photon energy, increases with $\Gamma$. When the requirement of transparency bounds the  comoving photon energy from above, this provides an upper limit on $\Gamma$. 
    
     A viewing offset of 10 or 15 degrees is not consistent with any and all models. We find, for example, that a viewing offset of 0.2 radians is incompatible with the prompt emission being powered by the same pool of kinetic energy that powers the long-term afterglow (e.g. internal shocks tapping about 1/2 of this kinetic energy for prompt emission and leaving the other half for  powering the afterglow),  for then, in order to power the afterglow, the accompanying electrons would be optically thick to the prompt photons unless their Lorenz factor was above several hundred.  But such a high Lorentz factor would give negligible emission at a viewing angle of at least 0.2 radians. The once popular internal shock model for prompt emission is thus challenged unless the opening angle for the observed prompt emission is higher than that of the afterglow core {\bf  (e.g. as in a structured jet or shock breakout model)} in which case the direction of the motion of the shocked material can be nearly along our line of sight. 
      
      If the baryonic material that powers the afterglow {\it also} powers the prompt emission, then we expect the opening angle of each to be more or less the same. On the other hand, if the material merely {\it reflects} the prompt emission, then there is not reason to expect that the latter need scale with the mass of the baryonic material, and the profile of the prompt emission could in fact be wider than the core of the afterglow.
    
   { Future GW signals from NS mergers may be viewed at even larger angles from the rotational axis. In this case the prompt emission would be even softer and dimmer than for GRB 170817A. In this case wide angle X-ray cameras would be the more suitable means of detecting an accompanying electromagnetic signal (Eichler and Guetta, 2010).}    \bigskip
    
    \bigskip
    
     \bigskip
     
    \centerline{\bf Acknowledgements}
    
    I thank Dr. N. Globus,  A. Levinson, R. Moharana and  E. Nakar for helpful conversations This work was supported by funding from the Israel Science Foundation, the Israel-U.S. Binational Science Foundation, and the Joan and Robert Arnow Chair of Theoretical Astrophysics.
    \bigskip

     \bigskip 
     \bigskip
    \bigskip
    
    \centerline{\bf References}
  
 \noindent Alexander, K.D.  et al. 2018, arXiv.org > astro-ph > arXiv:1805.02870
 
 \noindent Eichler, D. 2017,  ApJ. 85, 32
      
 \noindent Eichler, D.. 2005. ApJ,  628L, 17
 
 \noindent Eichler, D. 2014, ApJ, 787, L32
 
 \noindent Eichler, D. and Guetta, D. 2010, ApJ,  712, 392
 
 \noindent Eichler,, D., Guetta, D., and Manis, M. ApJ, 690, 61L
 
\noindent Eichler, D. and Jontof-Hutter, D. 2005,  ApJ, 635, 1182 

\noindent Eichler, D. and Manis, H. 2007, ApJ,  669. 65

\noindent Eichler, D. and Manis, H.,  2008, ApJ, 689L
 
 \noindent Fong, W., 2016, ApJ, 833, 151 
 
 \noindent Ioka, K.  and Nakamura,T.,  2018, arxiv.org/pdf/1710.05905
 
 \noindent Kaneko, Y. , Bostanci, Z. F., Gogus, E. and Lin, L.  2015,
 
 \noindent Lamb, G.P. Mandel, I, and Resmi, L. 2018, arXiv.org > astro-ph > arXiv:1806.03843
 
 \noindent Lazzati, D. et. al, 2018,  arXiv.org > astro-ph > arXiv:1712.03237
 
 \noindent Levinson, A. and Eichler, D. 1993, ApJ, 418, 386L
 
 \noindent Lyman, J.D. et al., 2018, arxiv.org/abs/1801.02669
 
 \noindent Margutti, R. et al., 2018, arXiv.org > astro-ph > arXiv:1801.03531
 
 \noindent Mooley, et al. 2017, arXiv.org > astro-ph > arXiv:1711.11573

 \noindent Mooley, K.P.,  et al., 2018,  arXiv.org > astro-ph > arXiv:1806.09693
  
 
 \noindent Nynka, M. , Ruan, J.J., Haggard, D.,, and Evans, P.A. , 2018, arXiv.org > astro-ph > arXiv:1805.04093

 \noindent Rees, M.J.  and Meszaros, P., 1992,  MNRAS, 258, 41P
 
 \noindent Resmi, L., et al. 2018, arXiv.org > astro-ph > arXiv:1803.02768
 
 \noindent Svensson, R., 1984, M.N.R.A.S.  209, 175 

 \noindent  Troja, E. et.al., 2018, arXiv.org > astro-ph > arXiv:1808.06617
 
 \noindent Veres, P.  et al.  2018,  astro-ph arXiv180207328V 
 
 \noindent Zhang,  B.-B.  et al. 2018, arXiv.org > astro-ph > arXiv:1712.03237

\end{document}